\documentclass[ amsmath, amssymb, superscriptaddress, aps, prx, reprint, floatfix, ]{revtex4-2}

\usepackage[version=4]{mhchem}
\usepackage{amsmath}
\usepackage{amssymb}
\usepackage{glossaries}
\usepackage{graphicx}
\usepackage{tabularx}
\usepackage{siunitx}
\usepackage{hyperref}
\usepackage[T1]{fontenc}
\usepackage{lmodern}

\hypersetup{
    unicode,
    colorlinks=true,
    urlcolor=black,
    linkcolor=black,
    citecolor=black
}

\graphicspath{{./figures/}}

\newcommand{\hmn}[1]{
  \ensuremath{\begingroup\setupHMN #1\endgroup}%
}

\newcommand{\setupHMN}{%
  \doHMN{-}{\HMNoverline}%
  \doHMN{*}{\HMNminverse}%
  \doHMN{i}{\infty}
}

\newcommand{\doHMN}[2]{%
  \begingroup\lccode`~=`#1
  \lowercase{\endgroup\let~}#2%
  \mathcode`#1="8000
}

\newcommand{\HMNminverse}[1]{\frac{#1}{m}}
\newcommand{\HMNoverline}[1]{\mkern1mu\overline{\mkern-1mu#1\mkern-1mu}\mkern1mu}

\setacronymstyle{long-short}
\newacronym{cx}{vdW-DF-cx}{van-der-Waals-density functional with consistent exchange}
\newacronym{ce}{CE}{cluster expansion}
\newacronym{dft}{DFT}{density functional theory}
\newacronym{led}{LED}{light-emitting diode}
\newacronym{mc}{MC}{Monte Carlo}
\newacronym{md}{MD}{molecular dynamics}
\newacronym{mlip}{MLIP}{machine-learned interatomic potential}
\newacronym{nep}{NEP}{neuroevolution potential}
\newacronym{sro}{SRO}{short-range order}
\newacronym{vcsgc}{VCSGC}{variance-constrained semi-grand canonical}

\DeclareSIUnit\angstrom{\text{Å}}
\DeclareSIUnit\atom{\text{atom}}
\DeclareSIUnit\atompercent{\text{at.}\unit{\percent}}

\makeatletter
\let\oldtheequation\theequation
\renewcommand\tagform@[1]{\maketag@@@{\ignorespaces#1\unskip\@@italiccorr}}
\renewcommand\theequation{(\oldtheequation)}
\makeatother

\global\let\oldnewlabel\newlabel
\gdef\newlabel#1#2{\newlabelxx{#1}#2}
\gdef\newlabelxx#1#2#3#4#5#6{\oldnewlabel{#1}{{#2}{#3}}}
\let\newlabel\oldnewlabel
\begin{document}

\title{Strong Impact of Halide Ordering on \texorpdfstring{\\}{}Structural Phase Transitions in Mixed Perovskites}

\newcommand{\addchalmers}{%
    Department of Physics and Astronomy, Chalmers University of Technology, SE-412 96 Gothenburg, Sweden
}
\newcommand{\addbirmingham}{%
    School of Chemistry, University of Birmingham, B15 2TT, Birmingham, UK
}

\author{Felix Uddén}
\affiliation{\addchalmers}
\author{Erik Fransson}
\affiliation{\addchalmers}
\author{Julia Wiktor}
\affiliation{\addchalmers}
\author{Benjamin M. Gallant}
\affiliation{\addbirmingham}
\author{Dominik J. Kubicki}
\affiliation{\addbirmingham}
\author{Paul Erhart}
\email{erhart@chalmers.se}
\affiliation{\addchalmers}


\begin{abstract}
Mixed halide perovskites are highly versatile semiconductors with applications in photovoltaics, light-emitting diodes, and photodetectors.
Understanding their thermodynamic phase behavior is central to guiding compositional design and improving device stability.
Here, we train \glspl{mlip} on \acrlong{dft} reference data for \ce{Cs_{x}Rb_{1-x}PbBr_{3y}I_{3-3y}}, \ce{Cs_{x}Rb_{1-x}PbBr_{3y}Cl_{3-3y}}, and \ce{Cs_{x}Rb_{1-x}PbCl_{3y}I_{3-3y}} halide perovskites, enabling large-scale hybrid \acrlong{mc}-\acrlong{md} simulations that sample both configurational and vibrational degrees of freedom.
All three binary halide systems exhibit a miscibility gap, the extent of which correlates with halide ion size mismatch. 
The gaps in Br--Cl and Br--I close at low temperatures, while the Cl--I gap extends above room temperature.
At temperatures above the miscibility gap (\qtyrange{200}{500}{\kelvin}), all systems show a tendency toward layered halide ordering, with halide species preferentially occupying apical or equatorial octahedral sites.
In \ce{CsPbBr_{3y}I_{3-3y}}, this ordering occurs in a device-relevant temperature regime and is linked to the structural phase transitions, shifting transition temperatures by up to \qty{100}{\kelvin} relative to randomly mixed structures.
We attribute the strongly non-linear composition dependence of the orthorhombic-tetragonal phase boundary observed experimentally (a linear decrease followed by a plateau) to halide ordering.
Introducing Rb on the A-site weakens halide ordering and eliminates the non-linear behavior, while narrowing the miscibility gap in both the Br--I and Br--Cl systems.
These results establish halide ordering as a key determinant of structural phase stability in mixed-halide perovskites.
\end{abstract}

\glsresetall

\maketitle

\section{Introduction}
Multi-component halide perovskites are versatile materials, finding applications in optoelectronic devices as diverse as photovoltaics \cite{wang2023suppressed}, \glspl{led} \cite{KonSunZha24}, photodetectors \cite{SakTurMat23}, and thermoelectrics \cite{XieHaoBao20}.
A substantial part of this success is due to the compositional tunability of the perovskite \ce{ABX3} structure, which enables optimization of key properties for optoelectronic performance \cite{LimCzLin22} and phase stability \cite{wang2023suppressed}.
Ion mixing on even a single site can significantly increase the complexity of a system through phase segregation and local structural variations, and correspondingly increase the complexity of the resulting structure--property relationships. 
Achieving and maintaining phase homogeneity is a critical challenge for perovskite optoelectronics \cite{wang2023suppressed, SheGalHol23, HokSloDoh15}, as there is increasing understanding of the role local structure and dynamics play in determining optoelectronic properties \cite{DubNeiKla25, OthJeaJac24, DohNagKub21}.
To maximize device performance and stability, perovskite compositions such as the Cs-rich mixed-halide perovskites used in multi-junction photovoltaics \cite{wang2023suppressed} and \glspl{led} \cite{YuaDaiSun24, CheWanKus25} often utilize simultaneous ion mixing across at least two of the three perovskite sites, further increasing the complexity.

Gaining insight into the thermodynamic behavior of mixed perovskites is therefore central to understanding their stability and guiding compositional design, particularly the interplay between structural phase transitions and compositional ordering.
Experimentally, bulk phase transition and segregation behavior in halide perovskites have been studied using, e.g., time-resolved optical emission spectroscopy \cite{wang2023suppressed, HokSloDoh15}, magnetic resonance \cite{wang2023suppressed, DutFraHai25}, diffraction measurements \cite{SheGalHol23}, and thermal analysis \cite{SimBalWil20, MozMauChe17}. 
Examining local phase segregation and the dynamic processes that drive such behavior is more challenging experimentally, but recent work using both neutron scattering \cite{DubNeiKla25, SwaStoPar15, DruPinRud16} and magnetic resonance methods \cite{SenMouKim17, gruninger2021microscopic} as atomic-level probes has shed light on composition-dependent processes including ion diffusion and formation of reduced-symmetry nanodomains.
Although powerful, such experimental methods are time-consuming and can typically only be conducted effectively on model systems for a limited number of compositions. 
Computational approaches offer a complementary perspective, enabling systematic exploration across the full composition space and allowing individual contributions, such as vibrational entropy and local structural distortions, to be isolated in a way that is challenging experimentally.

\Glspl{ce} parametrized against \gls{dft} have been used to study halide mixing in the Br--I, Cl--I, and Cl--Br systems \cite{Vanderven2018, YinYanWei14}, with one study additionally sampling chemical ordering via \gls{mc} simulations \cite{Vanderven2018}.
Both studies were restricted to a rigid lattice, and the results are therefore only reliable for the low-temperature orthorhombic phase.
\Gls{md} simulations using a reactive force field \cite{pols2023bromine} and a \gls{mlip} based on the MACE architecture \cite{LiaKlaWal25} have been applied to \ce{CsPbBr_{3y}I_{3-3y}} with random halide occupations, with the latter also exploring customized segregation patterns.

Here, we construct a \gls{mlip} based on the \gls{nep} formalism and trained on \gls{dft} reference data for compositionally mixed \ce{Cs_{x}Rb_{1-x}PbBr_{3y}I_{3-3y}}, \ce{Cs_{x}Rb_{1-x}PbBr_{3y}Cl_{3-3y}}, and \ce{Cs_{x}Rb_{1-x}PbCl_{3y}I_{3-3y}} halide perovskites.
The \gls{nep} formalism yields a strictly local and computationally efficient model, which is critical for the large number of energy evaluations required by \gls{mc} sampling and an advantage over graph neural network architectures that introduce explicit long-range interactions at substantially higher cost.
We use this framework in large-scale combined \gls{mc} and \gls{md} simulations \cite{SadErhStu12} that, for the first time in this type of system, sample both configurational and vibrational degrees of freedom to investigate composition-dependent phase transitions, ordering phenomena, octahedral tilting, and mixing energies as a function of temperature.

We find that in \ce{CsPbBr_{3y}I_{3-3y}}, halide ordering occurs in a device-relevant temperature regime and substantially suppresses the orthorhombic-tetragonal transition temperature by up to \qty{100}{\kelvin} relative to a randomly mixed structure.
Simulations with random occupation predict a linear variation in transition temperature across the full composition range, whereas fully equilibrated structures reproduce this non-linear behavior but with transition temperatures suppressed even further than observed experimentally. 
The experimental orthorhombic-tetragonal transition temperature lies between these two limits, consistent with an intermediate level of halide ordering.
In contrast, the cubic-tetragonal transition temperature is reproduced well by both random and ordered structures, indicating that this transition is largely insensitive to halide ordering.
This behavior in \ce{CsPbBr_{3y}I_{3-3y}} reflects broader trends: outside the miscibility gap, all three binary halide systems exhibit a tendency toward layered ordering, with different halides preferentially occupying either the apical or equatorial sites, and this ordering impacts the lower orthorhombic-tetragonal transition.


\section{Methods}

\begin{figure*}[t]
    \centering
    \includegraphics[]{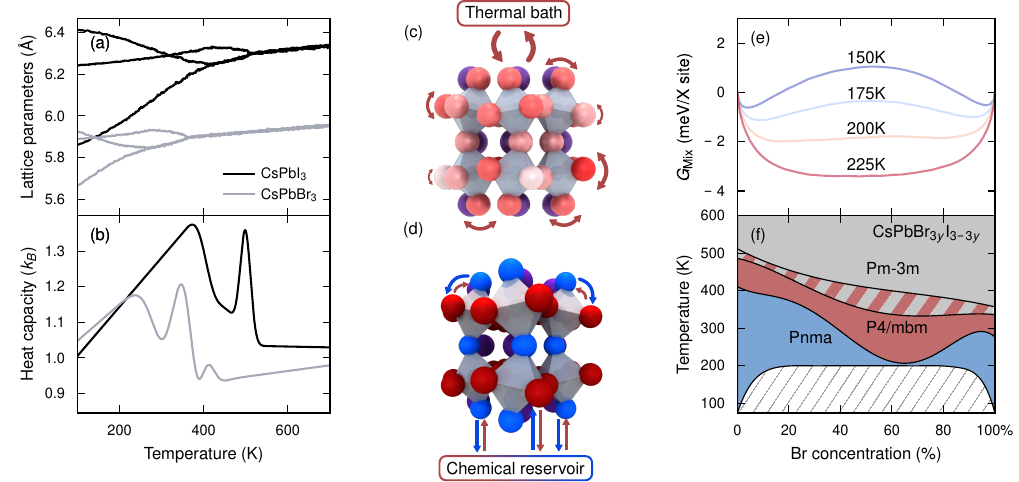}
    \caption{\textbf{Simulation approach.} 
    (a) Lattice parameters extracted for \ce{CsPbBr3} and \ce{CsPbI3} with \gls{md} demonstrating the three different structural phases with three, two, and one distinct lattice parameters corresponding to \hmn{Pnma}, \hmn{P4/mbm}, and \hmn{Pm-3m}, respectively. 
    (b) Heat capacity ($C_P$) for the same simulation. 
    (c) Vibrational degrees of freedom in (a) and (b), with the system coupled to an external heat bath. 
    (d) Chemical degrees of freedom simulated via \gls{mc} sampling in the canonical and \gls{vcsgc} ensembles. 
    (e) Gibbs free energy of mixing extracted by \gls{mc}\gls{md} sampling at different temperatures. 
    (f) Final phase diagram with structural transitions determined as in (a) and (b), and the miscibility gap determined from the free energy of mixing in (e).
    The red and gray striped region in (f) shows the hysteresis in the tetragonal-cubic transition temperature between the heating and cooling simulations.
    }
    \label{fig:simulation-approach}
\end{figure*}

\subsection{Density functional theory}
For all training structures, the energies, forces, and virial stresses were calculated using \gls{dft} as implemented in the Vienna ab-initio simulation package \cite{KreHaf93, KreFur1996-1, KreFur1996-2}.
The projector-augmented wave method \cite{Blo94, KreJou99} was used with a plane-wave energy cutoff of \qty{520}{\electronvolt}, using a Monkhorst--Pack mesh with a maximum k-point spacing of \qty{0.19}{\per\angstrom}.
For the exchange-correlation functional, we employed the \gls{cx} \cite{DioRydSch04, BerHyl2014}.

\subsection{Machine-learned interatomic potentials}
We constructed a \gls{mlip} based on the \gls{nep} framework \cite{FanZenZha21, FanWanYin22} for \ce{Cs_xRb_{1-x}PbCl_{3-3y-3z}Br_{3y}I_{3z}} following the iterative training strategy outlined in Ref. \citenum{FraWikErh23}. 
The training set comprised mixed-composition unit cells and supercells in the \hmn{Pnma}, \hmn{P4/mbm}, and \hmn{Pm-3m} phases, encompassing rattled and strained structures, \gls{md} snapshots, and phase-separated structures, i.e., structures with halide segregation.
Structural preparation, \gls{md} input generation, and post-processing were performed using the \textsc{ase} \cite{Larsen2017}, \textsc{calorine} \cite{Lindgren2024}, and \textsc{ovito} \cite{Stukowski2010} packages.

\Gls{md} structures were generated using an initial \gls{nep} model and selected for \gls{dft} labeling based on the uncertainty of their predicted energies, estimated from the predictions of an ensemble of models \cite{FraWikErh23}.
The final \gls{nep} model was then trained on the complete training set, using \textsc{gpumd} \cite{FanWanYin22}, comprising the initial reference structures and all newly selected \gls{md} snapshots.
A MACE model \cite{BatIlyKov22}, which is more accurate but substantially more expensive computationally \cite{LiaXuLin26}, was trained on the same dataset to validate the halide-ordering effects predicted by the faster \gls{nep} model.

Further validation of the \gls{nep} and MACE models is provided in the Supporting Information, including parity plots (\autoref{sfig:parity-plots}), the octahedral tilt angles (\autoref{stab:tilt-angles}), and comparison of lattice parameters to experiment \cite{HeStHa21, StMaPe13, MaRoBo18} (\autoref{sfig:lattice-parameters}).
The \gls{nep} and MACE models generated in this study are publicly available via Zenodo at \url{https://doi.org/10.5281/zenodo.22640311}.

\subsection{Molecular dynamics}
All \gls{md} simulations were carried out using the \textsc{gpumd} package \cite{XuBuPan25}.
Temperature and pressure were controlled using stochastic velocity rescaling \cite{BusDonPar07} and stochastic cell rescaling \cite{BerBus20}, respectively, with a timestep of \qty{5}{\femto\second} and at zero pressure.

Structural phase transition temperatures were determined from heating and cooling \gls{md} simulations, using a rate of temperature change in the range of \qtyrange{10}{14}{\kelvin\per\nano\second}, sufficiently slow to capture the relevant structural transitions \cite{FraWikErh23}.
The structural phase of each \gls{md} snapshot was identified by projecting onto the $M_3^+$ and $R_4^+$ modes of the cubic structure, which correspond to in-phase and out-of-phase octahedral tilting, respectively \cite{StoKisHat02, FraRosEri23}.
This was complemented by evaluation of the lattice parameters (\autoref{fig:simulation-approach}a) and heat capacity ($C_P$; \autoref{fig:simulation-approach}b), which exhibits a peak at both first-order and continuous phase transitions and was computed as
\begin{align*}
    C_P = \frac{\partial U}{\partial T}\approx\frac{\Delta U}{\Delta T},
\end{align*}
where $U$ is the potential energy.
To obtain a numerically stable estimate, the potential energy data were smoothed using a Hamming window of approximately \qty{0.5}{\kelvin}, after which the data were fitted with a piecewise polynomial of three segments connected at the transition temperatures, enabling better characterization of first-order transitions that are difficult to resolve from smoothed data alone.
Numerical differentiation of the resulting fit then yielded $C_P$ as a function of temperature.
Note that first-order transitions generally cannot be sampled using heating and cooling simulations. In this specific case, however, it is possible because the latent heat is extremely small \cite{FraRahWik23}.

\subsection{Monte Carlo--Molecular dynamics}
To account for both configurational and vibrational degrees of freedom (\autoref{fig:simulation-approach}c and d), a combined \gls{mc} and \gls{md} approach was employed.
Simulations alternated between \gls{mc} and \gls{md} segments, with each \gls{mc} segment consisting of one \gls{mc} cycle, i.e., $N$ \gls{mc} trial moves, where $N$ is the number of atoms included in the sampling; in the heating and cooling simulations $N=\num{61440}$.
The length of each \gls{md} segment was adjusted as a function of temperature to maintain an approximately constant number of accepted \gls{mc} trials per kelvin, with one \gls{mc} cycle every \num{500} \gls{md} steps at the lowest temperature.
In the case of \ce{CsPbBr_{3y}I_{3-3y}}, a partially equilibrated variant was also investigated, in which the setup was identical to the equilibrated case except that the \gls{mc} sampling rate was held constant (one \gls{mc} cycle every \num{5000} \gls{md} steps).
To determine the miscibility gaps, we performed \gls{mc} sampling in the \gls{vcsgc} ensemble, which couples the system to a chemical reservoir \cite{Fre23, SadErh12}.
While the constant-volume \gls{mc} trials sample the Helmholtz free energy, the interspersed NPT \gls{md} segments also include cell fluctuations, so that the combined scheme effectively samples the Gibbs free energy across the full concentration range of the relevant species, including regions of non-ideal mixing.

In the \gls{vcsgc} ensemble the relation between concentration and free energy is given by
\begin{align*}
    \frac{\partial G}{\partial c_i} = -2 k_B T \kappa \left(\frac{\phi_i - 2}{2} + 2 c_i \right),
\end{align*}
where $G$ is the Gibbs free energy per atom, $c_i$ is the concentration of species $i$, $\phi_i$ and $\kappa$ are parameters that constrain the concentration mean and variance, respectively, and $T$ is the simulation temperature.
We write this constraint in the form implemented in \textsc{gpumd}, which differs from the original \gls{vcsgc} formulation \cite{SadErh12} and the \textsc{icet}\cite{AngMunRah19} implementation:
\textsc{gpumd} adopts a symmetric convention in which $\phi_i = 0$ corresponds to $c_i = 0.5$ for a symmetric free energy landscape, rather than $\phi_i = -1$.
We point this out to avoid ambiguity when comparing the values of $\phi_i$ and $\kappa$ across different implementations.
The acceptance probability for each \gls{mc} trial is given by the Metropolis criterion, modified by an additional term in the exponent $\Delta E\rightarrow\Delta E + k_B T \kappa \left(\Delta\phi_i +2 \Delta c_i + 1/N \right)$.
When sampling the miscibility gap, a system size of \qty{25920}{\atom} was used.

For miscibility gap sampling, one \gls{mc} cycle was performed every \num{500} \gls{md} steps, allowing the system to equilibrate and accommodate the new configuration before the next \gls{mc} cycle. 
Simulations were run for \num{100000} \gls{md} steps (\qty{0.5}{\nano\second}), resulting in a total of \num{200} \gls{mc} cycles, ensuring convergence (\autoref{sfig:vcsgc-convergence}).
For systems with mixing on both A (cation) and X (halide) sublattices, \gls{mc} moves were alternated between sublattices, with the active sublattice switched every \num{4000} \gls{md} steps.

For simulations at fixed composition (i.e., heating and cooling runs), \gls{mc}\gls{md} sampling was performed in the canonical ensemble.
Each heating or cooling run spanned \qtyrange{1}{700}{\kelvin} over \qty{50}{\nano\second}, corresponding to \num{10000000} \gls{md} steps. 
Each canonical \gls{mc} trial consisted of swapping two atoms on the same sublattice, accepted according to the Metropolis criterion. 
Initially, these systems were equilibrated at the starting temperature for \num{2000000} \gls{md} steps (\qty{10}{\nano\second}), with a \gls{mc} cycle every \num{500} \gls{md} steps.
In heating simulations, one \gls{mc} cycle was performed every \num{500} \gls{md} steps at the lowest temperature and increasing logarithmically with temperature to \num{10000} \gls{md} steps at \qty{700}{\kelvin}.
For systems with mixing on both A and X sublattices, the active sublattice was alternated every \num{10000} \gls{md} steps, and the sampling rate was constant with \num{5000} \gls{md} steps between every cycle.

Concentrations throughout the paper are defined with respect to the sublattice of interest (e.g., in \ce{Cs_{0.8}Rb_{0.2}PbBr_{1.5}I_{1.5}} the Rb and Br site percentages are \qty{20}{\atompercent} on the cation sublattice and \qty{50}{\atompercent} on the halide sublattice, respectively).

\subsection{Phase diagram construction}
To construct the phase diagram from the Gibbs free energy obtained from the \gls{mc}\gls{md} simulations, we computed the free energy of mixing, defined as
\begin{align*}
    G_\text{mix}(x,T) = G(x,T) - \left( (1-x)G(0,T) + xG(1,T) \right),
\end{align*}
where $x$ is the fraction of the considered species, $T$ the temperature, and $G(x,T)$ the Gibbs free energy obtained from the \gls{mc}\gls{md} sampling described above.
Using a convex-hull construction, regions of excess free energy were identified from gaps in the convex hull, which indicate a miscibility gap.
By repeating this procedure at multiple temperatures (\autoref{fig:simulation-approach}e), the complete miscibility gap was determined.
Finally, combining the miscibility gap boundaries with the structural transition temperatures from heating and cooling runs, the full phase diagram was constructed (\autoref{fig:simulation-approach}f).

Alternatively, the free energy of mixing can be expressed using the Bragg--Williams approximation \cite{BraWil35}, in which the \qty{0}{\kelvin} mixing energy is combined with the ideal configurational entropy of a randomly mixed system,
\begin{align*}
    F_\text{mix}(x,T) &= U_\text{mix}(x,0) \\
    &\quad+ k_B T\left(x\ln(x) + (1-x)\ln(1-x)\right),
\end{align*}
where $U_\text{mix}(x,0)$ is the mixing energy at \qty{0}{\kelvin} for random mixing and $k_B$ is the Boltzmann constant.

\subsection{Structural order parameters}
As a further measure for determining the extent to which atoms on the two sublattices mix or segregate, we used the Warren--Cowley \gls{sro} parameter \cite{Cowley1950}, defined for an atom $i$ of type A as
\begin{align*}
    \alpha_i = 1 - \frac{Z_\text{B}}{Z_\text{tot}\, x_\text{B}},
\end{align*}
where $Z_\text{B}$ is the number of B neighbors in the first neighbor shell, $Z_\text{tot}$ is the total number of first-shell neighbors, and $x_\text{B}$ is the total fraction of B atoms in the structure.

In the orthorhombic (\hmn{Pnma}) and tetragonal (\hmn{P4/mbm}) phases, the reduced symmetry compared to the cubic (\hmn{Pm-3m}) phase allows layered halide ordering along the crystallographic $c$ direction, where the B-site atoms form layers separating the X sites into distinct apical and equatorial positions (\autoref{fig:simulation-approach}d).
To quantify the degree of layering, we used a second order parameter, defined analogously to Ref. \citenum{Vanderven2018} but with an additional normalization factor,
\begin{align*}
    \eta = \frac{x_\text{eq} - x_\text{ap}}{\eta_\text{max}},
\end{align*}
where $x_\text{eq}$ and $x_\text{ap}$ are the compositions on the equatorial and apical halide sites, respectively, and $\eta_\text{max}$ is the value of the unnormalized order parameter for complete segregation of one species to a single site type and can be expressed as
\begin{align*}
    \eta_\text{max} = \frac{3}{2} x  \left( 1 - \Theta\left(x-\frac{2}{3}\right) \right) + 3(1-x)\Theta\left(x-\frac{2}{3}\right),
\end{align*}
where $\Theta$ is the Heaviside function and $x$ the total fraction of the same species as in $\eta$.

\section{Results}
\begin{figure*}
    \centering
    \includegraphics{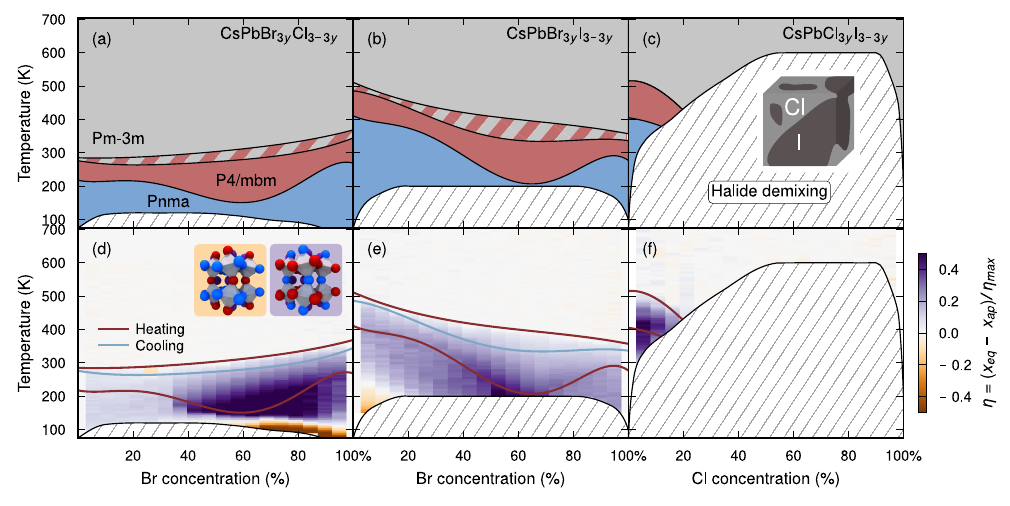}
    \caption{
    \textbf{Full phase diagrams for (a,d) \ce{CsPbBr_{3y}Cl_{3-3y}}, (b,e) \ce{CsPbBr_{3y}I_{3-3y}}, and (c, f) \ce{CsPbCl_{3y}I_{3-3y}}}.
    In (a--c), the three colored regions indicate the \hmn{Pnma} (blue), \hmn{P4/mbm} (red), and \hmn{Pm-3m} (gray) phases.
    The red and gray striped region shows the hysteresis in the tetragonal-cubic transition temperature between the heating and cooling simulations.
    White dashed regions indicate miscibility gaps, as shown in the inset in (c).
    (d--f) Layered ordering parameter and structural phase transition temperatures from heating and cooling runs.
    Insets in (d) show the positive and negative layered ordering extremes.
    }
    \label{fig:full-phase-diagrams}
\end{figure*}

\subsection{Phase diagrams}
The phase diagrams of all binary halide mixtures were determined using the combined \gls{mc}\gls{md} approach (\autoref{fig:full-phase-diagrams}a--c).
First, in \ce{CsPbBr_{3y}Cl_{3-3y}} (\autoref{fig:full-phase-diagrams}a) we identify the cubic \hmn{Pm-3m} phase for all compositions at temperatures above \qty{350}{\kelvin}. 
At lower temperatures (\qtyrange{150}{350}{\kelvin}), only the $M_3^+(z)$ mode is active, corresponding to the tetragonal \hmn{P4/mbm} phase.
Finally, at temperatures below \qty{250}{\kelvin}, the $M_3^+(z)$, $R_4^+(x)$, and $R_4^+(y)$ modes all become active, indicating the structure is in the orthorhombic \hmn{Pnma} phase.
At temperatures below \qty{125}{\kelvin}, there is a halide miscibility gap.

Second, in \ce{CsPbBr_{3y}I_{3-3y}} (\autoref{fig:full-phase-diagrams}b), both structural transitions are shifted to higher temperatures than in Br--Cl, consistent with the higher transition temperatures of \ce{CsPbI3} compared to \ce{CsPbCl3}. 
The \hmn{P4/mbm}-\hmn{Pm-3m} transition occurs between \qtyrange{350}{500}{\kelvin}, depending on halide composition, and the \hmn{Pnma}-\hmn{P4/mbm} is in the region of \qtyrange{200}{410}{\kelvin}.
As in \ce{CsPbBr_{3y}Cl_{3-3y}}, there is a miscibility gap. 
However, this gap closes at a higher temperature, approximately \qty{200}{\kelvin}.

Finally, in \ce{CsPbCl_{3y}I_{3-3y}} (\autoref{fig:full-phase-diagrams}c), only small regions of \hmn{Pnma} and \hmn{P4/mbm} are present, and only for Cl concentrations below \qty{25}{\atompercent}, with transition temperatures of \qty{400}{\kelvin} and \qtyrange{420}{510}{\kelvin} for the lower and upper transitions.
Comparing the extent of the halide miscibility gap at room temperature with experiment, neither \ce{CsPbBr_{3y}Cl_{3-3y}} nor \ce{CsPbBr_{3y}I_{3-3y}} exhibit any segregation \cite{JuYuYa23} in agreement with our simulations.  
For \ce{CsPbCl_{3y}I_{3-3y}} the extent of the miscibility gap indicated by our simulations at \qty{300}{\kelvin} aligns well with the experimentally determined boundaries, where mixing occurs only for Cl concentrations below \qty{15}{\atompercent} Cl. 
Moreover, the transition temperatures of both the Br--Cl and Br--I systems in the orthorhombic (\hmn{Pnma}) structure agree well with previous simulations using \glspl{ce} \cite{Vanderven2018}.
In contrast, restricting the comparison to the orthorhombic phase, the Cl--I system exhibits a miscibility gap extending to \qty{600}{\kelvin}, considerably higher than \qty{410}{\kelvin} reported in Ref. \citenum{Vanderven2018}.  
This discrepancy may stem from the inclusion of vibrational degrees of freedom in our \gls{mc}\gls{md} simulations, which capture the corresponding entropy contribution that is absent in a \gls{ce} approach and raises the predicted critical temperature in this system.
Additionally, the \gls{ce} model limited to an orthorhombic structure while the \gls{mc}\gls{md} approach allows for structural phase transitions, which is relevant in this case as the structure transitions to a cubic phase at these elevated temperatures.
To evaluate this effect, we compare against two \gls{ce} models trained on the \gls{nep} and MACE mixing energies, the construction of which is detailed in the Supporting Information. 
The resulting \gls{ce} critical temperatures, \qty{500}{\kelvin} and \qty{550}{\kelvin}, are closer to the previously reported \qty{410}{\kelvin} than the \gls{nep} \gls{mc}\gls{md} result, although full agreement is not achieved (\autoref{sfig:cli-miscibility-gap-comparison}).

\subsection{Halide anion ordering}
Our simulations reveal pronounced layered ordering (\autoref{fig:full-phase-diagrams}d inset) in \ce{CsPbBr_{3y}I_{3-3y}} and \ce{CsPbBr_{3y}Cl_{3-3y}}, outside the halide miscibility gap in both systems (\autoref{fig:full-phase-diagrams}d and e).
The layering order parameter ($\eta$) demonstrates a clear preference for Br at the apical sites in Br--Cl mixtures (\autoref{fig:full-phase-diagrams}d) and equatorial sites in Br--I mixtures (\autoref{fig:full-phase-diagrams}e).
The degree of layering reaches maximum $\eta$ values of 0.8 and 0.5 at approximately \qty{74}{\atompercent} Br and \qty{53}{\atompercent} Br for the Br--Cl and Br--I mixtures respectively, corresponding to a \qty{31}{\atompercent} and \qty{42}{\atompercent} difference in Br concentration between the apical and equatorial sites.
Such layered ordering has previously been reported, experimentally and computationally, for both organic \cite{BrCaWa16, OvLeGr22, HopCorMis24} and inorganic \cite{Vanderven2018, LiYuHe18} lead-based Br--I perovskites.
In contrast to previous studies, we investigate both chemical mixing and structural dynamics simultaneously and thus can directly disentangle their effect on each other.
The strongest layered ordering effect coincides with the transition from \hmn{Pnma}-\hmn{P4/mbm}, in the range \qtyrange{35}{90}{\atompercent} Br in the Br--Cl system and \qtyrange{40}{70}{\atompercent} Br for Br--I.
No layered ordering is observed in the cubic \hmn{Pm-3m} phase, consistent with its higher symmetry.
In \ce{CsPbCl_{3y}I_{3-3y}} only low levels of ordering are present, for Cl concentrations below \qty{25}{\atompercent} and temperatures below \qty{500}{\kelvin}, where $\eta$ is approximately 0.4.

In heating runs, $\eta$ shows the same sign across the entire compositional range in all systems (i.e., Br prefers the apical sites). 
In contrast, cooling runs of both Br--Cl and Br--I systems lead to $\eta$ with both signs, indicating either negative or positive ordering (\autoref{fig:full-phase-diagrams}d inset).
This suggests that positive ordering is energetically favorable compared to negative ordering, but that the difference is small. 
For example, the difference is only \qty{2.1}{\milli\electronvolt\per\atom} in the \hmn{Pnma} phase of the Br--I system for perfectly ordered structures with \qty{67}{\atompercent} Br.

\begin{figure*}
    \centering
    \includegraphics{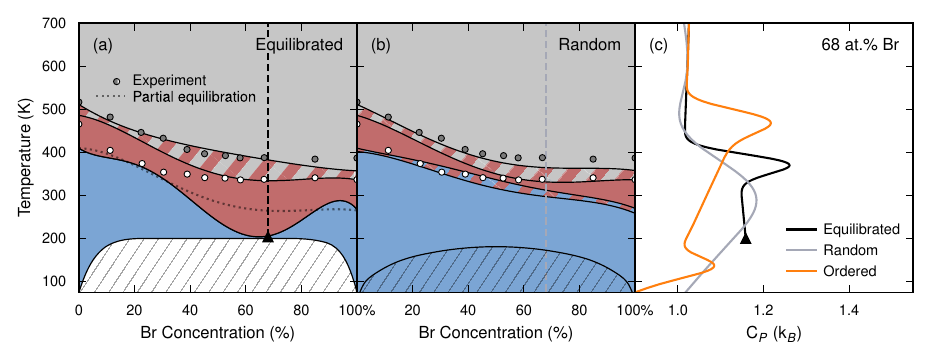}
    \caption{
    \textbf{Random and ordered halide perovskites.}
    Phase diagram of \ce{CsPbBr_{3y}I_{3-3y}} for equilibrated (a) and random (b) halide configurations.
    (c) Heat capacity for random, equilibrated, and ordered structures with \qty{68}{\atompercent} Br (random and equilibrated illustrated as dashed lines in a and b).
    In (a) and (b), the three colored regions indicate the \hmn{Pnma} (blue), \hmn{P4/mbm} (red), and \hmn{Pm-3m} (gray) phases.
    The red and gray striped region shows the region where heating and cooling simulations give different transition temperatures.
    Dashed regions indicate miscibility gaps, determined using \gls{vcsgc} in (a) and the Bragg--Williams approximation in (b).
    Experimental data in (a) and (b) are from \citeauthor{NaBeMa20} \cite{NaBeMa20}.
    The dotted line presented in (a) is the \hmn{Pnma}-\hmn{P4/mbm} transition temperature for a partially equilibrated system.
    }
    \label{fig:random-and-ordered-halides}
\end{figure*}

In \ce{CsPbBr_{3y}I_{3-3y}}, the regions of strongest ordering and of the \hmn{Pnma}-\hmn{P4/mbm} transition both occur near room temperature, making this ordering phenomenon relevant to typical optoelectronic device operating conditions.
The ordering itself involves local rearrangement of halide atoms between apical and equatorial sites, a process that requires sufficient halide mobility.
By contrast, in \ce{CsPbBr_{3y}Cl_{3-3y}} the analogous region lies near \qty{150}{\kelvin}, well below device-relevant temperatures.
Therefore, we use \ce{CsPbBr_{3y}I_{3-3y}} as a model system. 
To assess the impact of ordering on phase transitions, we compare the experimental transition temperatures with simulations performed using random and equilibrated configurations (\autoref{fig:random-and-ordered-halides}a and b) \cite{NaBeMa20}. 
For the transition temperature from \hmn{Pnma}-\hmn{P4/mbm}, experimental results show strongly non-linear composition dependence of the orthorhombic-tetragonal phase boundary: a linear decrease from approximately \qty{460}{\kelvin} to \qty{340}{\kelvin} between \qty{0}{\atompercent} and \qty{50}{\atompercent}, and a plateau in temperature above \qty{50}{\atompercent} Br.
This trend suggests a suppression in transition temperatures at intermediate Br compositions.
Similarly suppressed transition temperatures have been observed experimentally in other Br--I perovskites, for example in \ce{MAPbBr_{3y}I_{3-3y}} at high Br concentrations \cite{LeFrTo19}.
This non-linear composition dependence of the orthorhombic-tetragonal transition is observed in the equilibrated configuration, but the suppression effect is stronger than that observed experimentally (\autoref{fig:random-and-ordered-halides}a). 
By contrast, the random configuration exhibits an approximately linear dependence on Br concentration and is thus appears inconsistent with the experimentally observed trend (\autoref{fig:random-and-ordered-halides}b).
We therefore suggest that the experimentally observed trend could be due to slow kinetics, leading to only partially halide-ordered states (\autoref{fig:random-and-ordered-halides}a) intermediate between the random and fully ordered limits.
The partially ordered structure presented here clearly captures the lower phase transition better than both the random and fully equilibrated structures.

For the higher-temperature \hmn{P4/mbm}-\hmn{Pm-3m} transition, ordering is limited even in the equilibrated structures.
As a result, the transition temperature is only slightly altered in the equilibrated configuration relative to the random one and both adequately reproduce the experimental trend.

Although the transition temperature for the  \hmn{P4/mbm}-\hmn{Pm-3m} transition is nearly identical between the equilibrated and random structures, there is a clear difference when considering the heat capacity(\autoref{fig:random-and-ordered-halides}c).
Focusing on \qty{68}{\atompercent} Br, for the random configuration there is only a single broad peak for both transitions.
In the equilibrated case, we see a clear peak for the second \hmn{P4/mbm}-\hmn{Pm-3m} transition but none for the lower \hmn{Pnma}-\hmn{P4/mbm} transition.
This is because the lower transition lies inside the halide miscibility gap. 
When instead considering a structure with fixed chemical ordering, there are clear peaks in the heat capacity both transitions.

Ordering also affects the extent of the miscibility gap.
The miscibility gaps determined from \gls{mc}\gls{md} simulations (representing full equilibration with corresponding ordering) and from \gls{md} simulations of fully random structures differ in extent.
The gap for the random structures closes \qty{25}{\kelvin} lower and spans a larger range of halide compositions than that obtained with \gls{vcsgc} (\autoref{fig:random-and-ordered-halides}).
There are multiple possible contributions to this effect.
First, the Bragg--Williams approximation can overestimate the entropic term, as it describes a transition directly from a segregated to a completely random state rather than through the partially ordered regime we observe outside the miscibility gap.
Second, the Bragg--Williams approximation excludes contributions from different phases. However, in this case the different phases have only a small impact due to their second-order character.
Third, vibrational entropy is also excluded from the approximation.
What this result demonstrates is the importance of correctly modeling and considering chemical ordering, phase transitions, and phonon behavior in order to make quantitative predictions for these systems.

\begin{figure}
    \centering
    \includegraphics{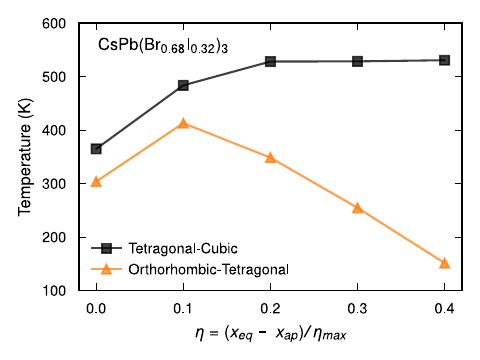}
    \caption{
    \textbf{Tunable phase transition temperatures.}
    Structural phase transition temperatures as a function of the halide ordering level for \ce{CsPbBr_{3y}I_{3-3y}} at \qty{68}{\atompercent} Br.
    The \hmn{Pnma}-\hmn{P4/mbm} transition (orange triangles) and the \hmn{P4/mbm}-\hmn{Pm-3m} transition (black squares) are shifted in the ranges \qtyrange{200}{400}{\kelvin} and \qtyrange{375}{525}{\kelvin}, respectively, by systematically varying the degree of halide ordering.
    }
    \label{fig:tunable-transition-temperatures}
\end{figure}

Next, to investigate the effect of different levels of ordering, we create structures with $\eta$ between 0 and 0.4 run heated simulations from \qty{1}{\kelvin} (\autoref{fig:tunable-transition-temperatures}). 
When fixing the level of ordering, the \hmn{Pnma}-\hmn{P4/mbm} transition temperature shifts in the range \qtyrange{200}{400}{\kelvin}.
Note that in these simulations the chemical ordering is constant and thus the upper transition is also impacted by ordering, with the \hmn{P4/mbm}-\hmn{Pm-3m} transition shifting from \qty{375}{\kelvin} to \qty{525}{\kelvin}.
Experimentally, such a setup would be difficult to reproduce, as the system prefers a random configuration at the higher temperatures and is not miscible at the lower temperatures. 
However, these results demonstrate that there is a close connection between chemical ordering and the soft phonon $M_3^+$ and $R_4^+$ modes governing the phase transitions in these mixed-halide perovskites.
This indicates that, to model phase transitions in these materials, the inclusion of both chemical ordering and vibrational effects is crucial.

\begin{figure*}[t]
    \centering
    \includegraphics{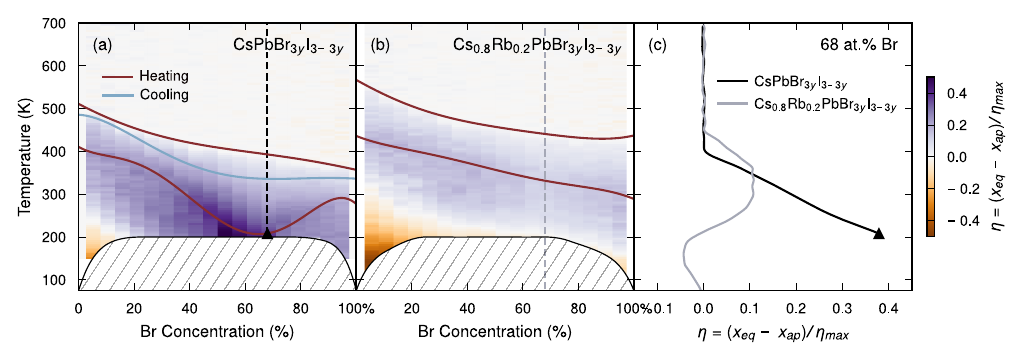}
    \caption{
    \textbf{Halide ordering with Rb.}
    Layered ordering parameter and structural phase transition temperatures from heating and cooling runs for \ce{CsPbBr_{3y}I_{3-3y}} (a) and \ce{Cs_{0.8}Rb_{0.2}PbBr_{3y}I_{3-3y}} (b).
    (c) Layered ordering parameter for both compositions at \qty{68}{\atompercent} Br (dashed lines in a and b).
    }
    \label{fig:halide-ordering-with-rb}
\end{figure*}

\subsection{A-site alloying with Rb}
To assess the effect of A-site alloying, we replace \qty{20}{\atompercent} of Cs with Rb in \ce{Cs_{0.8}Rb_{0.2}PbBr_{3y}I_{3-3y}}.
Introducing Rb raises both the \hmn{Pnma}-\hmn{P4/mbm} and \hmn{P4/mbm}-\hmn{Pm-3m} transition temperatures by \qtyrange{10}{125}{\kelvin} across the entire composition range (\autoref{fig:halide-ordering-with-rb}b).
For the \hmn{P4/mbm}-\hmn{Pm-3m} transition, the temperature increase reaches \qty{70}{\kelvin} at the pure boundary compositions but falls to only \qty{40}{\kelvin} at \qty{40}{\atompercent} Br.
Within the compositional range that exhibits strong layered ordering in the \ce{CsPbBr_{3y}I_{3-3y}} system, the \hmn{Pnma}-\hmn{P4/mbm} transition increases by \qty{125}{\kelvin}. 
This increase in transition temperature is consistent with Rb-based perovskites exhibiting larger octahedral tilt angles and greater stability in the \hmn{Pnma} and \hmn{P4/mbm} phases than their Cs-based counterparts (\autoref{stab:tilt-angles}).
The increased stability can be attributed to the smaller size of Rb compared to Cs, which results in shorter Pb--X distances and stronger bonding \cite{LiFuPe20}. 
Structurally, a smaller cation will increase tilting due to the larger A-site cavity being incompletely filled by the cations.
This reasoning can be formalized using the Goldschmidt tolerance factor \cite{Gol26},
\begin{align*}
    t= \frac{r_A+r_X}{\sqrt{2}(r_B+r_X)},
\end{align*}
where $r_A$, $r_B$, and $r_X$ are the ionic radii of A, B, and X atoms.
According to the Goldschmidt factor, the smaller ionic radius of Rb relative to Cs favors a more stable orthorhombic structure in which larger octahedral tilt angles can accommodate smaller ions, in agreement with our observations.

When Rb is introduced, the transition between \hmn{Pnma} and \hmn{P4/mbm} does not retain its non-linear dependence on halide composition.
Instead, the transition temperature varies linearly with Br concentration, closely resembling the Rb-free configuration in which halides are randomly mixed (\autoref{fig:random-and-ordered-halides}b).
Moreover, when Rb is introduced, halide ordering becomes substantially less pronounced, with $\eta$ decreasing to a maximum of approximately \num{0.1} (\autoref{fig:halide-ordering-with-rb}b). 
This maximum also occurs at lower Br concentration than in the Rb-free system, and at considerably higher temperatures. 
This comparison between the Rb-free and Rb-containing systems neatly highlights the strong correlation between atomic-level halide ordering and phase transition temperature suppression.
The weakening of ordering at elevated temperatures in the Rb-containing system can be attributed to the increasing role of configurational entropy, which favors a more disordered state.

Notably, the introduction of Rb makes \gls{sro} on the A-site sublattice possible (\autoref{sfig:short-range-order}c).
For \ce{Cs_{0.8}Rb_{0.2}PbBr_{3y}I_{3-3y}} with $<$\qty{10}{\atompercent} or $>$\qty{90}{\atompercent} Br, the magnitude of A-site \gls{sro} increases to approximately \num{0.15}.
At temperatures above \qty{200}{\kelvin}, the \gls{sro} decreases below \num{0.05} across the full compositional range.
Rb also introduces a region with oppositely layered halide ordering in the Br--I system (\autoref{fig:halide-ordering-with-rb}b), at temperatures and compositions below \qty{210}{\kelvin} and \qty{30}{\atompercent} Br.
This ordering overlaps with the region of increased affinity of Br for Rb (\autoref{sfig:short-range-order}a).

\subsection{Halide miscibility with Rb}
Ionic radii are a key property for the miscibility of different halide species due to the contribution of local strain to the free energy.
A large radius mismatch is expected to inhibit mixing, as observed in our results.
\ce{CsPbCl_{3y}I_{3-3y}} exhibits the largest halide size mismatch and correspondingly shows the highest miscibility-gap critical temperature.
In contrast, \ce{CsPbBr_{3y}Cl_{3-3y}} has the smallest radius mismatch between its pure phases and the lowest critical temperature.
\ce{CsPbBr_{3y}I_{3-3y}} lies between these two systems in terms of the mismatch and accordingly displays a critical temperature between the previous two.

Introducing \qty{20}{\atompercent} Rb on the A-site reduces the extent of the miscibility gaps for the mixed Br--Cl and Br--I systems but not for the Cl--I alloy (\autoref{sfig:halide-miscibility-gaps}). 
For Br--I and Br--Cl, this reduction is most pronounced in the compositional range where there is negative Rb--X \gls{sro}.
In the same region, the Rb--Cs \gls{sro} tendency remains weak, indicating a preference for an almost random distribution of Cs and Rb on the A-site and thus also increasing X-site mixing.

The decreased extent of the miscibility gap could possibly be explained with the following two effects.
First, the additional mixing entropy stemming from the A-site sublattice could lead to an overall larger entropic contribution for the mixed halide state compared to the phase separated state, leading to a smaller more narrow halide miscibility gap.
Second, as argued in Ref.~\citenum{YinYanWei14}, is that the smaller radius of Rb and thus shorter distances between halide sites would increase the energetic gain of mixing.
Due to the difference in electronegativity between halide species, the charges are redistributed, preferentially to the halide with higher electronegativity, thus resulting in a more stable configuration.
Thus, the extent of the miscibility gap for Br--I and Br--Cl can be attributed to two possible effects: correlated A- and X-site mixing that increases the configurational entropy, and smaller distances between halide atoms due to the smaller Rb ionic radius.


\section{Conclusion}
Our results demonstrate that all three binary halide perovskite compositions studied exhibit pronounced atomic-level halide ordering, even when halides are globally well mixed, with layered structures appearing in which halides preferentially occupy either apical or equatorial sites.
In \ce{CsPbBr_{3y}I_{3-3y}}, this ordering occurs at device-relevant temperatures and is closely coupled to the soft cubic phase $M_3^+$ and $R_4^+$ phonon modes that govern transitions between orthorhombic, tetragonal, and cubic phases.
Further, layering impacts the soft phonon dynamics far above the transition temperature, which is crucial for electron-phonon coupling \cite{HaiFraDut25}.
We attribute the suppression of the orthorhombic-tetragonal phase boundary at intermediate Br compositions to layered halide ordering, but find that the reduction in transition temperature is larger in our fully equilibrated simulations than observed experimentally. 
This points to the role of kinetics in experimental samples, which can inhibit equilibration and leave them in a partially ordered state.

By systematically varying the degree of layering in our simulations, we can tune both the orthorhombic-tetragonal and tetragonal-cubic phase transitions, shifting them by up to \qty{100}{\kelvin}.
These results demonstrate that halide ordering governs the stability of structural phases in mixed-halide perovskites, with direct consequences for the performance and optoelectronic properties of photovoltaic devices.
At lower temperatures, all binary halide compositions exhibit a halide miscibility gap. 
The compositional range over which this gap is observed is proportional to the ionic radius mismatch between the component halides.
While the miscibility gaps of both the Br--Cl and Br--I systems close well below room temperature (\qty{125}{\kelvin} and \qty{200}{\kelvin}, respectively), the gap observed in the Cl--I system persists up to \qty{600}{\kelvin}.
Introducing \qty{20}{\atompercent} Rb on the A-sublattice suppresses the miscibility gap in both Br--Cl and Br--I systems, a suppression we suggest is may be caused by correlated A- and X-site mixing that increases the difference in entropy between segregated and mixed states, as well as lattice contraction due to the smaller Rb ionic radius.
In conclusion, these results highlight the central role of both halide ordering and phase separation in determining the structural behavior of mixed-halide perovskites.

\section*{Acknowledgments}
Funding from the Knut and Alice Wallenberg Foundation (No. 2024.0042) and the Chalmers Academic Excellence Program is gratefully acknowledged.
The computations were enabled by resources provided by the National Academic Infrastructure for Supercomputing in Sweden (NAISS) at \url{https://www.naiss.se/} partially funded by the Swedish Research Council through grant agreement no. 2022-06725, as well as the Berzelius resource provided by the Knut and Alice Wallenberg Foundation at NSC.
B.M.G. and D.J.K. acknowledge the UKRI Horizon Europe guarantee funding (PhotoPeroNMR, Grant Agreement EP/Y01376X/1).

The authors used AI tools (Claude, Anthropic, and ChatGPT, OpenAI) during the preparation of this manuscript for assistance with writing and language improvements.
All scientific content, interpretation, and final wording were reviewed and approved by the authors.

\section*{Supporting Information}

The Supporting Information provides details pertaining to model training and validation.

\section*{Data Availability}

The \gls{nep} models constructed in this study as well as the data presented in this paper is openly available via Zenodo at \url{https://doi.org/10.5281/zenodo.22640311}.

\bibliography{references.bib}

\end{document}